\documentclass{article}
\usepackage{spconf}
\usepackage{cite}
\usepackage{amsmath,amssymb,amsfonts}
\usepackage{algorithmic}
\usepackage{graphicx}
\usepackage{textcomp}
\usepackage{array}
\usepackage{multirow}
\usepackage{xcolor,mdframed}
\usepackage{stfloats}
\usepackage{booktabs}
\usepackage{grffile}
\usepackage{hyperref}
\usepackage[skip=0pt]{caption}
\usepackage{soul}
\usepackage{tikz}
\usepackage{enumitem}
\setlist{nosep, leftmargin=14pt}
\usetikzlibrary{calc}

\makeatletter
\newif\if@anonymize

\@anonymizefalse  % Uncomment to show text

\if@anonymize
\newcommand{\highlight@DoHighlight}{
	\fill [outer sep = -15pt, inner sep = 0pt, color=black]
	($(begin highlight)+(0,8pt)$) rectangle ($(end highlight)+(0,-3pt)$) ;
}

\newcommand{\highlight@BeginHighlight}{
	\coordinate (begin highlight) at (0,0) ;
}

\newcommand{\highlight@EndHighlight}{
	\coordinate (end highlight) at (0,0) ;
}

\newdimen\highlight@previous
\newdimen\highlight@current
\newlength{\item@width}

\DeclareRobustCommand*\anonymize{%
	\SOUL@setup
	\def\SOUL@preamble{%
		\begin{tikzpicture}[overlay, remember picture]
		\highlight@BeginHighlight
		\highlight@EndHighlight
		\end{tikzpicture}%
	}%
	\def\SOUL@postamble{%
		\begin{tikzpicture}[overlay, remember picture]
		\highlight@EndHighlight
		\highlight@DoHighlight
		\end{tikzpicture}%
	}%
	\def\SOUL@everyhyphen{%
		\discretionary{%
			\SOUL@setkern\SOUL@hyphkern
			\SOUL@sethyphenchar
			\tikz[overlay, remember picture] \highlight@EndHighlight ;%
		}{%
		}{%
			\SOUL@setkern\SOUL@charkern
		}%
	}%
	\def\SOUL@everyexhyphen##1{%
		\SOUL@setkern\SOUL@hyphkern
		\settowidth{\item@width}{##1}%
		\makebox[\item@width]{}%
		\discretionary{%
			\tikz[overlay, remember picture] \highlight@EndHighlight ;%
		}{%
		}{%
			\SOUL@setkern\SOUL@charkern
		}%
	}%
	\def\SOUL@everysyllable{%
		\begin{tikzpicture}[overlay, remember picture]
		\path let \p0 = (begin highlight), \p1 = (0,0) in \pgfextra
		\global\highlight@previous=\y0
		\global\highlight@current =\y1
		\endpgfextra (0,0) ;
		\ifdim\highlight@current < \highlight@previous
		\highlight@DoHighlight
		\highlight@BeginHighlight
		\fi
		\end{tikzpicture}%
		\settowidth{\item@width}{\the\SOUL@syllable}%
		\makebox[\item@width]{}%
		\tikz[overlay, remember picture] \highlight@EndHighlight ;%
	}%
	\SOUL@
}
\else
\newcommand{\anonymize}[1]{#1}
\fi
\makeatother

\newcommand{\comment}[1]{}

\renewcommand{\b}[1]{\mathbf{#1}}

\def\argmin{\mathop{\mathrm{arg\,min}}} % car l'indice est reparti

\title{Hybrid learning of Non-Cartesian k-space trajectory and MR image reconstruction networks}
\name{\anonymize{Chaithya G R$^{\star \dagger}$} \qquad \anonymize{Zaccharie RAMZI$^{\star \dagger \ddagger}$} \qquad \anonymize{Philippe CIUCIU$^{\star \dagger}$}}

\address{\anonymize{$^{\star}$CEA, Joliot, NeuroSpin, Universit\'e Paris-Saclay, F-91191 Gif-sur-Yvette, France} \\
	\anonymize{$^{\dagger}$Inria, Parietal, Universit\'e Paris-Saclay, F-91120 Palaiseau, France} \\
	\anonymize{$^{\ddagger}$AIM, CEA, CNRS, Universit\'e Paris-Saclay, Universit\'e Paris Diderot}}
\begin{document}
\ninept

\maketitle
\begin{abstract}
Compressed sensing (CS) in Magnetic resonance Imaging (MRI) essentially involves the optimization of 1) the sampling pattern in k-space under MR hardware constraints and 2) image reconstruction from the undersampled k-space data. Recently, deep learning methods have allowed the community to address both problems simultaneously, especially in the non-Cartesian acquisition setting. This paper aims to contribute to this field by tackling some major concerns in existing approaches.
Regarding the learning of the sampling pattern, we perform ablation studies using parameter-free reconstructions like the density compensated (DCp) adjoint operator of the nonuniform fast Fourier transform~(NUFFT) to ensure that the learned k-space trajectories actually sample the center of k-space densely. Additionally we optimize these trajectories by embedding a projected gradient descent algorithm over the hardware MR constraints. Later, we introduce a novel hybrid learning approach that operates across multiple resolutions to jointly optimize the reconstruction network and the k-space trajectory and present improved image reconstruction quality at 20-fold acceleration factor on $T_1$ and $T_2$-weighted images on the fastMRI dataset with SSIM scores of nearly 0.92-0.95 in our retrospective studies.
\end{abstract}

\section{Introduction}

Speeding up magnetic resonance imaging~(MRI) acquisitions involves jointly optimizing how the k-space is undersampled and how image quality is preserved (or aliasing artifacts are discarded) during image reconstruction. Based on the CS theory, the k-space has to be undersampled according to a variable density sampling (VDS)~\cite{puy2011variable,chauffert2013variable, Chauffert_SIAM2014,adcock2013breaking,boyer2017compressed}, whose shape depends on the underlying anatomy. In 2D imaging, 2D VDS patterns can only be efficiently achieved by using non-Cartesian sampling.
Recently, SPARKLING~\cite{Lazarus_MRM_19,Lazarus_NMRB_20,3dsparkling} was introduced as an iterative procedure to optimize a non-Cartesian k-space sampling pattern according to a prescribed target sampling density~(TSD) with each k-space trajectory being compliant with MR hardware constraints~(particularly maximum gradient $G_\text{max}$ and slew rate $S_\text{max}$). Further, the algorithm results in locally uniform sampling patterns and thus avoids holes and clusters in k-space.
However, a major limitation of SPARKLING is the need for a TSD as an input in the optimization process. To address this issue, the TSD was learned in a deep learning setting in~\cite{learning_density} using LOUPE~\cite{loupe2020} as an acquisition model. 
While this method provided adequate reconstruction performances, there was still a mismatch in the learning process. Indeed, the gridded TSD was learned in the Cartesian domain, while the actual trajectory being optimized was non-Cartesian. As this could lead to suboptimal results, there is a need to jointly learn both the TSD and the reconstruction network in a non-Cartesian setting.
\begin{figure*}[h!]
	\includegraphics[width=\textwidth]{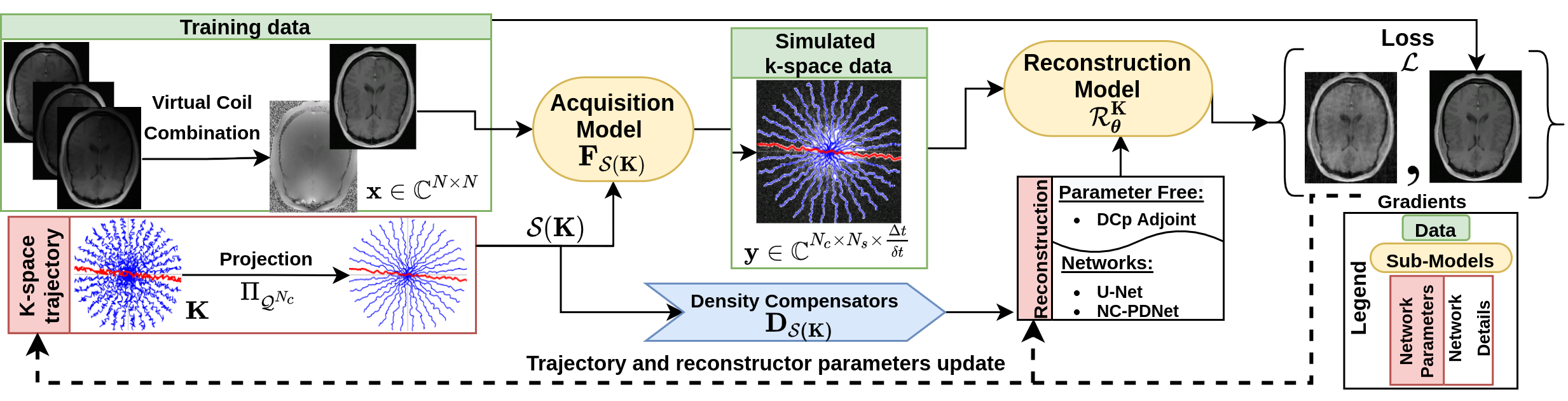}
	\caption{\label{fig:gen_model} A generic learning-based framework for joint optimization of the MRI acquisition and reconstruction models. }
\end{figure*}

More recent methods~\cite{pilot, bjork,3d-flat} bypass the need for estimating a TSD, by jointly learning the trajectories and the reconstruction network in a data-driven manner. In~\cite{pilot, 3d-flat}, the authors use a gridding step followed by a U-net to reconstruct the MR images and jointly learn the network parameters and k-space sampling trajectory. This method, however, relies on auto-differentiation of the NUFFT operator, which may not be very accurate~(see \cite{wang_ISMRM21}), thus resulting in suboptimal local minima. This was actually observed as the final learned trajectories only slightly deviated from their initialization. In \cite{bjork}, the authors use \cite{wang_ISMRM21} to obtain a more accurate Jacobian approximation of the NUFFT operator. Both above referenced approaches\cite{pilot, bjork} enforce the hardware constraints with the help of a penalty on the cost function. Although a viable option, this does not guarantee that the optimized trajectories will meet these constraints. Additionally this requires tuning the hyperparameter associated with this additional penalty in the cost function. Further, this penalty could affect the overall gradients, thereby resulting in suboptimality. To overcome these issues, the authors in \cite{bjork} parameterized the trajectory with B-spline curves, which reduce the number of inequality constraints and hence the influence of this penalty. However, such parameterization could severely limit the trajectories and prevent them from better exploring the k-space. Finally, both methods do not make use of density compensation~(DCp) which plays a crucial role in obtaining cleaner images in the non-Cartesian deep learning setting\cite{ncpdnet}.

Given these results, in this work, we first present a generic model for jointly optimizing the non-Cartesian trajectories and the corresponding reconstruction network. More precisely, we introduce a method to learn the trajectories in a data-driven manner while embedding a projected gradient descent algorithm\cite{Chauffert_TMI_16} to fulfill the hardware constraints.
Second, we show experimentally that a novel generali\-zed {\em hybrid learning} scheme performs better compared to a joint or alternated learning of the trajectory and the reconstruction network. Further we compare the trajectories obtained under different reconstruction settings~(DCp Adjoint, DCp Adjoint + U-Net and NC-PDNet) and evaluate their performances in terms of image quality metrics~(SSIM~\cite{ssim}, PSNR) in a retrospective evaluation framework.
% We introduce our learning strategy of trajectories that operates in a data-driven manner while embedding a projected gradient descent algorithm\cite{Chauffert_TMI_16} to fulfill the hardware constraints. 

\section{Generic Model}
In this section we provide details~(see Fig.~\ref{fig:gen_model}) about the different components of the general model that is used for optimizing the non-Cartesian sampling trajectories as well as the image reconstruction network.

\subsection{Acquisition}
	
\quad~~\textbf{Data:}
In this work, we relied on the brain k-space data from the fastMRI dataset\cite{zbontar2018fastmri} for learning the k-space trajectories. We applied the virtual coil method~\cite{parker2014phase} to combine per-channel images to obtain a complex-valued single channel image, which was used in our training stage. This was done to account for the phase accrual during k-space data acquisition and make the learned trajectories more realistic.
Following the formulation developed in~\cite{3dsparkling}, we denote an MR image $\mathbf{x} \in \mathbb{C}^{N \times N}$, over a field of view $\mathcal{F} \times \mathcal{F}$. Then the 2D k-space of the image is defined in $[-K_{\rm max}, K_{\rm max}]^2$, with $K_{\rm max} = \frac{N}{2 \mathcal{F}}$. In all our trajectories, we kept $N=320$ and $\mathcal{F}=0.23$ m. For the sake of simplicity, we normalize the k-space to $\Omega = [-1, 1]^2$.

\textbf{Trajectory Specification:}
In our work, we directly learn the k-space trajectory locations unlike BJORK \cite{bjork} where the coefficients of B-spline curves were learned. We believe that such parameterizations could potentially limit the exploration in the k-space.
Instead we optimize the 2D k-space sampling pattern $\mathbf{K}$ which is composed of $N_c$ shots, $\mathbf{K}=(\mathbf{k}_i)_{i=1}^{N_c}$. Each shot can be played by the scanner hardware at the pace of gradient raster time $\Delta t$, throughout the readout time $T_{obs}$, resulting in $N_s = \lfloor \dfrac{T_{\rm obs}}{\Delta t}\rfloor$ samples per shot. Thus the overall sampling pattern is $\mathbf{K} \in \Omega^{N_c \times N_s}$. Throughout this work, we used $N_s = 512$ ($T_{\rm obs} \approx 5\text{ms}$, for $T_1$-w and $T_2$-w imaging) and $N_c=16$, leading to an acceleration factor $AF = \frac{N}{N_c}=20$.

\textbf{Forward Model:}
For the sake of simplicity, here we model the MR acquisition process using the NUFFT modeled by $\mathbf{F_K}$ for the corresponding k-space sampling pattern $\mathbf{K}$. In practice, the k-space data is sampled at a $\delta t$-pace, with $\delta t$ the dwell time used by the analog to digital converter~(ADC). Hence a more realistic modeling of k-space data $\b{y} \in \mathbb{C}^{N_c \times N_s \times \frac{\Delta t}{\delta t}}$ is as follows:
%i.e. the sampling period 
\begin{align}\label{eq:forward_model}
	\b{y} &= \b{F}_{\mathcal{S}(\b{K})} \b{x} 
\end{align}
\noindent where $\mathcal{S}(\b{K}) \in \Omega^{N_c \times N_s \times \frac{\Delta t}{\delta t}}$ is the sampling pattern obtained after \emph{linearly} interpolating each shot in $\b{K}$ by a factor of $\frac{\Delta t}{\delta t}$. In this work we used $\delta t = 2$ \textmu s. This interpolation is done to model the ADC hardware in place at the scanner. Note that at a cost of increased complexity more realistic forward models can be considered to take off-resonance effects as well as $T_2^*$-decay into account. However, they were not included in this work. %$B_0$-inhomogeneities and $T_2^*$ parameter maps.

\subsection{Reconstruction}
Throughout this work, we denote the reconstruction network as $\mathcal{R}^\mathbf{K}_{\boldsymbol{\theta}}(\b{y})$, which is a parameterized by $\boldsymbol{\theta}$ and depends on the k-space trajectory $\mathbf{K}$ being optimized (through the interpolated trajectory $\mathcal{S}(\b{K})$) and its corresponding k-space measurements $\b{y}$.

\textbf{DCp Adjoint:}
As a first step, we rely on a simple DCp adjoint NUFFT to perform image reconstruction. This ensures that the given network actually results in trajectories which are data-aware and samples the center of k-space more densely as expected by traditional CS theory \cite{puy2011variable,chauffert2013variable, Chauffert_SIAM2014,adcock2013breaking,boyer2017compressed}.
Note that in this case the reconstruction network is parameter-free and satisfies the following reconstruction formula:
\begin{align}
	\widehat{\b{x}}_{dc} &= \mathcal{R}^\b{K}(\b{y}) = \b{F^H_{\mathcal{S}(\b{K})}} \b{D_{\mathcal{S}(\b{K})}} \b{y} 
\end{align}
\noindent where $\b{F^H}$ is the adjoint of NUFFT operator $\b{F}$ and $\b{D_{\mathcal{S}(\b{K})}}$ is the density compensators for the interpolated sampling pattern $\mathcal{S}(\b{K})$ obtained iteratively using the method described in\cite{pipe_dc}.

\textbf{U-Net:}
We then used a U-net architecture\cite{unet} directly on the density-compensated adjoint image $\widehat{\b{x}}_{dc}$. %of the k-space measurements $\b{y}$ obtained above
This network can be trained to reduce aliasing and other artifacts in the final reconstructed image.

\textbf{NC-PDNet:}
Finally, we made use of the state-of-the-art DCp unrolled network  called NC-PDNet\cite{ncpdnet}. This network is known to provide improved image quality compared to U-Net and remain robust to changes in trajectories, which is crucial in our approach as the sampling pattern is learned and updated in every gradient descent step. A key difference of this network as compared to that used in BJORK~\cite{bjork} is that we use DCp to better condition the reconstruction problem. Moreover, in the underlying CNN-based denoiser, we use different parameters across iterations, giving improved reconstruction performance at the cost of larger memory footprint.

\section{Optimization scheme}

We trained our networks for nearly 1.5k steps with batch size of 64 on the fastMRI training data, which was split into training and validation in a 90\%-10\% ratio. This enabled early stopping to prevent overfitting. The original fastMRI validation dataset was used only for later evaluation.

\subsection{Costs, gradients and optimizers}
We optimize for the k-space trajectory $\b{K}$ and reconstruction network $\mathcal{R}^\b{K}_\theta$ as follows:
\begin{align}\label{eq:global_cost}
	(\widehat{\b{K}}, \widehat{\boldsymbol{\theta}}) &= \argmin_{(\b{K} \in \mathcal{Q}^{N_c}, \boldsymbol{\theta})} {\cal L}\left(\b{x}, \mathcal{R}^\b{K}_{\boldsymbol{\theta}} \left( \b{F}_{\mathcal{S}(\b{K})} \b{x} \right) \right)
\end{align}
where the hardware constraints on the trajectories are specified by the feasible trajectory space $\mathcal{Q}^{N_c}$ as described in Sec.~\ref{sec:constraints}. The loss function  ${\cal L}$ used in this study was inspired by~\cite{pezzotti_loss} which is a weighted sum of $\ell_1$, $\ell_2$ and multi-scale structural similarity index~($\mathcal{S}$) \cite{mssim}:
\begin{align*}
	{\cal L}(\b{x}, \widehat{\b{x}}) &= \alpha (1 - \mathcal{S}(\b{x}, \widehat{\b{x}})) + \bar{\alpha} ||\b{x} - \widehat{\b{x}}||_1 + \frac{\bar{\alpha}^2}{2} ||\b{x} - \widehat{\b{x}}||_2
\end{align*}
with $\bar{\alpha}=1-\alpha$ and the value of $\alpha$ was tuned to 0.998 to give nearly equally balanced terms.

Optimizing the trajectory in~\eqref{eq:global_cost} requires computing the gradient of ${\cal L}$ with respect to $\b{K}$:
%would need the calculation of the following gradients:
\begin{align}
\label{eq:gradK}
\frac{\partial \mathcal{L}(\b{x}, \b{\widehat{x}})}{\partial \b{K}} &= \nabla \mathcal{L}(\b{x}, \b{\widehat{x}}) \frac{\partial  \b{\widehat{x}}}{\partial \b{K}} = \nabla \mathcal{L}(\b{x}, \b{\widehat{x}}) \frac{\partial \mathcal{R}^\mathbf{K}_{\boldsymbol{\theta}}(\b{y}) }{\partial \b{K}}
\end{align}
In order to simplify this gradient calculation and reducing its computational complexity, we neglect the contribution of gradients from density compensators $D_{\mathcal{S}(\mathbf{K})}$.
For the case of DCp Adjoint reconstruction, Eq.~\eqref{eq:gradK} reads:
\begin{align*}
\frac{\partial \mathcal{L}}{\partial \b{K}} &= \nabla\mathcal{L} \left(\frac{\partial  \b{\widehat{x}}}{\partial \b{D_{\mathcal{S}(\b{K})}}\b{y}} \b{D_{\mathcal{S}(\b{K})}} \frac{\partial \left(\b{F}_{\mathcal{S}(\b{K})}\b{x}\right)}{\partial \b{K}} + \frac{\partial \b{F}^H_{\mathcal{S}(\b{K})}}{\partial \b{K}} \right)
\end{align*}
In order to obtain the gradient of NUFFT operators $\b{F}_{\mathcal{S}(\b{K})}$ and $\b{F}^H_{\mathcal{S}(\b{K})}$ with respect to the k-space trajectory $\b{K}$, we implemented\cite{wang_ISMRM21} in \anonymize{tfkbNUFFT}. As these underlying gradients vary extremely in norm depending on the k-space region (as noted in \cite{degournay2021}), we used the ADAM optimizer for learning the trajectories, while we relied on rectified-ADAM for optimizing the image reconstruction network $\mathcal{R}^\b{K}_\theta$. We used a learning rate of $10^{-3}$.

\subsection{Multi-resolution}
\label{sec:multi_res}

Following\cite{lebrat_optimal_2019}, we used a multi-resolution approach to optimize the k-space trajectory and allow for the algorithm to reach faster convergence. Consequently, the sampling pattern was first optimized on down-sampled trajectories. After optimizing over $N_{ds} \approx 250$ steps, the solution $\widehat{\b{K}}^{n_1}$ at the resolution level $n_1$ was then interpolated and further used as a warm restart for the up-sampled problem $n_2=2\, n_1$. We used dyadic scaling and trained our trajectory over 6 decimation levels (i.e. $n_1=N_c\times N_s/2^5$ up to $n_6=2^5 n_1$). 
% thus starting at the 64-fold decimation.
At any decimation level $n_k$, the solution $\widehat{\b{K}}^{n_k}$ used in the forward model~\eqref{eq:forward_model} was still linearly interpolated as ${\cal S}(\widehat{\b{K}}^{n_k})$ to match the final trajectories shape and ensure that the reconstructor sees the same amount of data at every $n_k$. %~(i.e. same $N_s$)

\subsection{Learning schemes}

Here there are two sub-models that are being optimized: The k-space trajectory $\b{K}$ and the image reconstruction network $\mathcal{R}^\b{K}_\theta$. These sub-models can be trained in the following ways:
%Following PILOT\cite{pilot} and BJORK\cite{bjork},

\textbf{Joint Learning (JL):}
In an ideal scenario, we would like to jointly optimize both sub-models. This idea was pushed forward pretty successfully in both PILOT and BJORK\cite{pilot,bjork} at the cost of producing trajectories close to their initialization. Here, 
as we rely on an unconstrained parameterization of $\b{K}$ this scheme may lead to poor local minimizers. The reason is that the reconstruction network $\mathcal{R}^\b{K}_{\boldsymbol{\theta}}$ is initially naive, and thus not trained on various types of trajectories, hence it may give poor reconstructed images associated with high losses ${\cal L}$ for certain configurations of ${\cal S}(\b{K})$.
This initial bad conditioning of  $\mathcal{R}^\b{K}_{\boldsymbol{\theta}}$ cannot be addressed by updating $\b{K}$.
Consequently, the gradient of ${\cal L}$ with respect to $\b{K}$~(see~\eqref{eq:gradK}) is corrupted leading to trajectory updates in wrong directions. 
%by such high loss values.
%update of these bad losses pass through our network and update the trajectory to improve the reconstructor performance. 
% with a poor initial reconstructor by updating trajectory and thus these gradients lead the trajectory towards sub-optimality. 

\textbf{Alternative Descent (AD):}
In this scenario, we first optimize the trajectory using the DCp Adjoint reconstruction method. As the latter is parameter-free, the optimized sampling pattern $\widehat{\b{K}}$ is purely data-driven and it densely samples the center of k-space, hence providing reasonable image reconstruction. Later on, we learn $\mathcal{R}^{\widehat{\b{K}}}_\theta$. Although this method can provide improved results, it still does not truly learn the sub-models in a joint fashion.

\textbf{Hybrid Learning (HL):}
In this third scenario, we try to retain the best of both previous schemes, by slowly moving from AD to JL. As a first step, we start off by only learning the trajectory $\b{K}$ using DCp adjoint as reconstruction. Following this, we carry alternative descent between $\b{K}$ and $\mathcal{R}^\b{K}_\theta$ while slowly increasing the number of steps to move from noisy to more accurate gradients over time. In practice, this was done at different resolution levels as prescribed in Tab.~\ref{tab:hybrid_learning}.

\vspace{.3cm}
\begin{table}[h]
		\caption{Hybrid learning strategy \label{tab:hybrid_learning}}
	\begin{center}
	\footnotesize
		\begin{tabular}{cc}
			\toprule
			Decimation levels  &      Learning Method \\
			\midrule
			32 $\rightarrow$  4 &  Alternative descent for 20 steps each  \\
			4 $\rightarrow$  1 &   Alternative descent for $N_{ds}/2$ steps each\\
			1  &  True joint optimization throughout \\
			\bottomrule
		\end{tabular}
	\end{center}
\end{table}

\vspace{-1cm}
\subsection{Constraints and Projection}
\label{sec:constraints}
In order to obtain realistic learned trajectories which can be run by a scanner gradient hardware, we need to impose $G_\text{max}$ and $S_\text{max}$ constraints. Under the normalized sampling domain $\Omega = [-1, 1]^2$, we obtain the feasible constraint set $\mathcal{Q}^{N_c}$ from\cite[Eq.~(2)]{3dsparkling}. To enforce these constraints, we project all shots in $\b{K}$ using the projector $\Pi_{\mathcal{Q}^{N_c}}$ developed in\cite[Sec.~S1]{3dsparkling} after each gradient descent step (for more details, see \cite{Chauffert_TMI_16}).
With this in mind, given a step size $\boldsymbol{\eta}_t$,
each update of the trajectory along the learning process reads:
\begin{align}\label{eq:traj_update}
	\b{K}_{t+1} &\leftarrow \Pi_{\mathcal{Q}^{N_c}}\left(\b{K}_{t} - \boldsymbol{\eta}_t \frac{\partial \mathcal{L}(\b{x}, \b{\widehat{x}})}{\partial \b{K}} \right)
\end{align}

Given this projection step our trajectories strictly meet the constraints on $G_{\rm max}$ and $S_{\rm rmax}$ in contrast to earlier works\cite{pilot,3d-flat,bjork} which handle these constraints on the trajectories through a penalty term in the loss ${\cal L}$. Importantly, this projection step is parameter-free. Additionally, the projector has no influence over the gradient of ${\cal L}$ with respect to $\b{K}$, allowing it to improve image quality at the reconstruction stage.
%The use of a projector is particularly useful as it prevents tuning of hyper parameters associated with penalty weighting in the overall cost function. Also, a projector has zero influence over the gradient of the cost function with respect to trajectory, allowing it to best optimize for reconstruction image quality.

\section{Numerical studies}
We compare our results for $\text{AF}=20$, and 2 imaging contrasts~(T1-w, T2-w) using AD and HL as learning strategies. We did not compare with JL startegy as our experiments led to low SSIM scores~(between 0.7 and 0.8 after training). Further, for comparison with an earlier baseline, we use SPARKLING trajectories generated with the learned TSD using LOUPE as obtained in\cite{learning_density}.\footnote{Comparisons with PILOT \cite{pilot} and BJORK \cite{bjork} could not be done due to lack of time, although we thank the authors for sharing their trajectories with us. We will compare them in future work.}. We denote this approach by SP.
The learned models were tested on 512 slices from the fastMRI validation dataset. We used the SSIM and PSNR scores on each slice as image quality metrics for evaluation.

\begin{figure}[h]
	\includegraphics[width=\columnwidth]{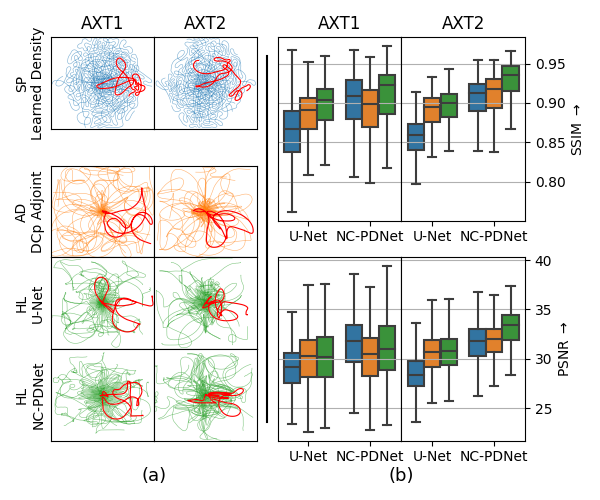}
	\caption{\label{fig:box_plots} {\bf (a)} The optimized k-space trajectories with SPARKLING on learned density (top, blue) and by joint optimization (bottom). For joint optimization, results with AD (orange) and HL (green) methods are presented for (AX)T1/(AX)T2 contrasts. Finally, trajectories optimized with U-Net and NC-PDNet as reconstructors shown for HL method. {\bf (b)} Box plots summarizing the corresponding color-matched image reconstruction results on a retrospective study using 512 slices~(fastMRI validation dataset). SSIMs/PSNRs appear on top/at the bottom.}
\end{figure}

\subsection{Quantitative}

We first present the trajectory and quantitative comparisons for each of the learning strategies in the form of box plots in Fig.~\ref{fig:box_plots}. One key aspect of all our learned trajectories is that they are extremely curvy (in contrast to \cite{pilot, bjork}) and maximally explore the k-space, which ideally will lead to better performance. We notice that for the AD strategy, where the trajectory was learned with parameter free DCp adjoint as reconstructor, we do sample the center of k-space densely as expected by CS theory.

We observe that NC-PDNet consistently outperforms U-Net as it is an unrolled network. Further, we see that HL learning strategy clearly outperforms all the other methods with SSIM scores in the range of 0.9-0.95. 
The only exception to this is PSNR values for $T_1$-w contrasts, where SP presents equivalent results to NC-PDNet. However, note that the SP trajectories were obtained by learning the density, followed by obtaining trajectories with SPARKLING and finally learning the NC-PDNet network for it. HL strategy on the other hand learns jointly in a single training.

\newcommand{\filenametone}{file\_brain\_AXT1\_202\_2020283.h5}
\newcommand{\filenamettwo}{file\_brain\_AXT2\_209\_6001069.h5}
\begin{figure}[h]
	\includegraphics[width=\columnwidth, trim={0.5cm, 0.8cm, 0.5cm, 0.3cm},clip]{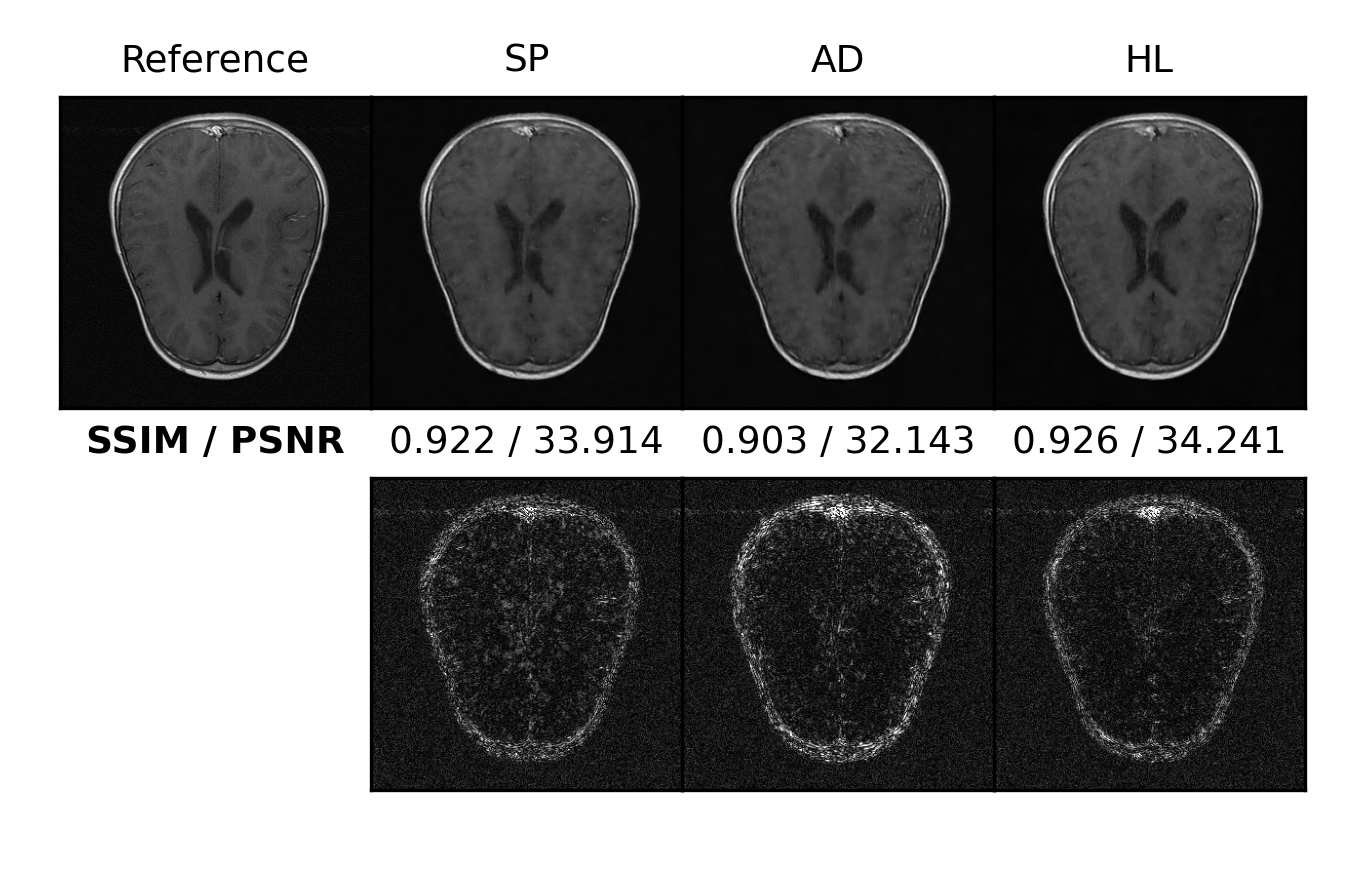}
	\caption{\label{fig:single_result_T1}\textbf{Top:} Reference and reconstruction results for a single slice from \texttt{\filenametone} with SP, AD and HL strategies and NC-PDNet reconstructor. \textbf{Bottom:} The residuals, scaled to match and compare across methods.}
\end{figure}

\begin{figure}[h]
	\includegraphics[width=\columnwidth, trim={0.5cm, 0.8cm, 0.5cm, 0.3cm},clip]{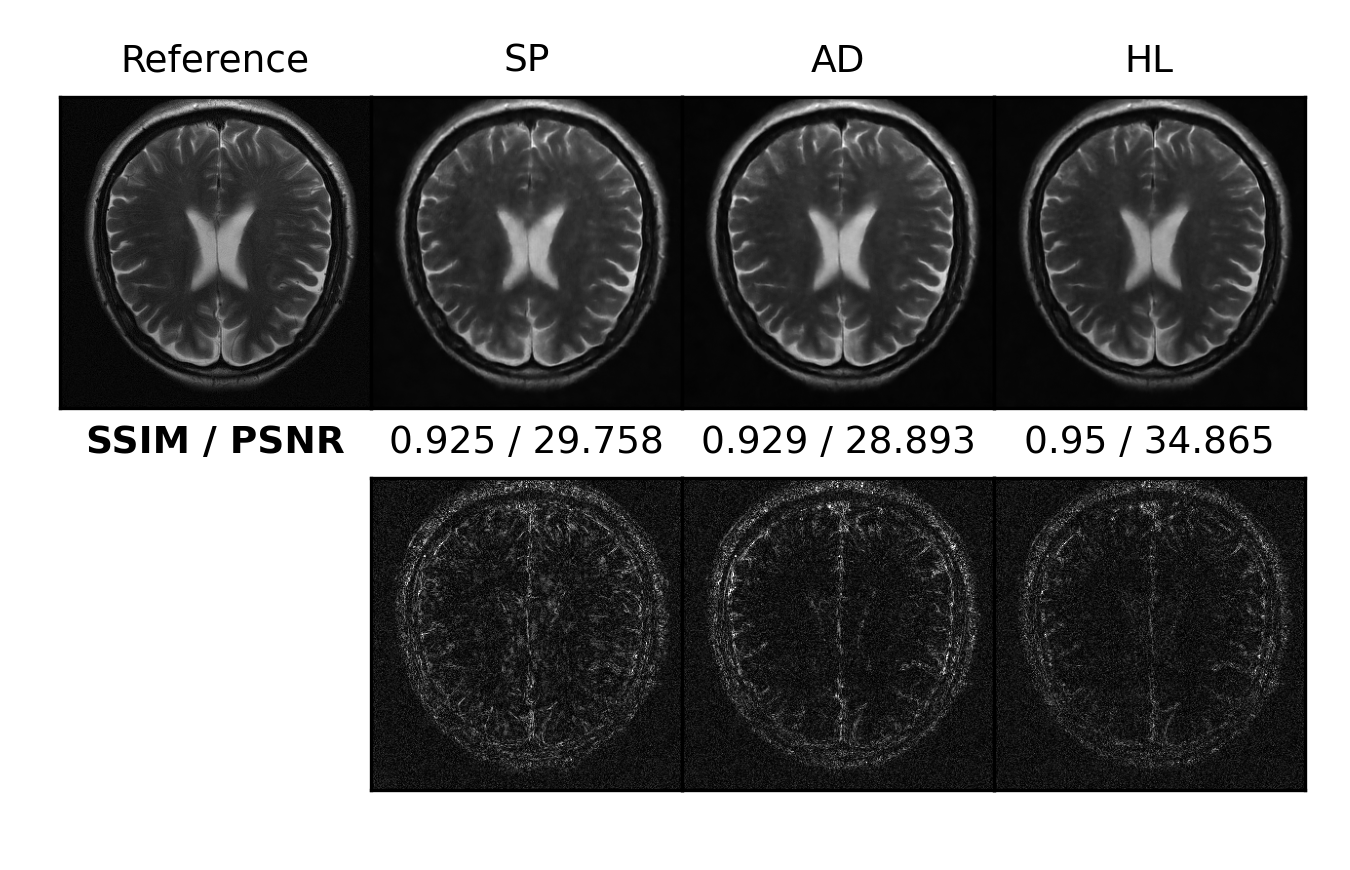}
	\caption{\label{fig:single_result_T2}\textbf{Top:} Reference and reconstruction results for a single slice from \texttt{\filenamettwo} with SP, AD and HL strategies and NC-PDNet reconstructor. \textbf{Bottom:} The residuals, scaled to match and compare across methods.}
\end{figure}

\subsection{Qualitative}
In order to assess image quality and the residuals across the different learning schemes~(SP, AD, HL), we present a single reconstructed slice for $T_1$-w images in Fig.~\ref{fig:single_result_T1} and $T_2$-w images in Fig.~\ref{fig:single_result_T2}. We clearly observe lesser residual structures in the HL/NC-PDNet approach, which can also be verified quantitatively with highest SSIM and PSNR scores for 0.926 and 34.241, respectively for $T_1$-w images. However, it is notable to observe that SP approach gives at par results as observed previously. For  $T_2$-w, we see HL/NC-PDNet approach clearly outperforms with SSIM and PSNR scores of 0.95 and 34.865 respectively.

\section{Conclusions}
In this work, we present a hybrid framework for jointly learning the trajectory and image reconstructor through projected gradient descent, which provides trajectories comparable to those generated by SPARKLING with learned density. Through restrospective studies on the fastMRI validation dataset, we showed that this novel learning scheme works across multiple resolutions and leads to superior performance of the trajectories and improved image quality
overall.

Future prospects of this work include prospective implementations through modifications of $T_1$ and $T_2$-w imaging sequences. This could lead to additional constraints on the trajectory arising from the MR sequencing or contrasts.

\section*{Acknowledgments}	
\anonymize{This work was granted access to the HPC resources of IDRIS under the allocation 2021-AD011011153 made by GENCI. Chaithya G R was supported by the CEA NUMERICS program, which has received funding from the European Union's Horizon 2020 research and innovation program under the Marie Sklodowska-Curie grant agreement No 800945.}
\bibliographystyle{IEEEtran}
\bibliography{ref.bib}

\end{document}